# Examining users' preferences towards vertical graphical toolbars in simple search and point tasks

RAFAL MICHALSKI†

†Institute of Organization and Management (I23),
Faculty of Computer Science and Management (W8)
Wrocław University of Technology,
27 Wybrzeże Wyspiańskiego,
50-370 Wrocław, POLAND
e-mail: Rafal.Michalski@pwr.wroc.pl

Corresponding author:

Dr. Rafał Michalski

Institute of Organization and Management (I23),
Faculty of Computer Science and Management (W8)
Wrocław University of Technology,
27 Wybrzeże Wyspiańskiego,
50-370 Wrocław, POLAND
e-mail: Rafal.Michalski@pwr.wroc.pl
phone: +48 71 348 50 50
fax: +48 71 320 34 32

## Abstract

The main purpose of this study was to investigate the nature of preferences and their relation to the objective measures in simple direct manipulation tasks involving both the cognitive process as well as the visually guided pointing activities. The conducted experiment was concerned with the graphical structures resembling toolbars widely used in graphical interfaces. The influence of the graphical panel location, panel configuration as well as the target size on the user task efficiency and subjects' preferences were examined. The participants were requested to express their attitudes towards the tested panels before and after the efficiency examination. This subjective evaluation was carried out within the framework of Analytic Hierarchy Process (AHP; Saaty, 1977; 1980). The subjective results that were obtained showed significant differences in the subjects' preferences towards examined panels before and after completing the tasks. It seems that the users are able to comparatively quickly change their minds after gaining some experience with the investigated stimuli. Additionally, the applied cluster analysis revealed that the subjects were not homogenous in their opinions, and they formed groups having similar preference structures.

## Keywords

## Introduction

The people's preferences play a fundamental role in almost any decision making process. For that reason they have been subject to a wide interest of many researches in multiple scientific areas. The understanding of the true nature of the individuals' likings seems also to be an important issue in the Human-Computer Interaction (HCI) domain. From the usability of graphical interfaces point of view, the users' preferences are directly connected with the user satisfaction which in turn is one of the main dimensions of the usability concepts defined both by HCI researchers and practitioners (Nielsen, 1993; Dix et al., 2004; Folmer and Bosch, 2004). This component can also be found in the usability international standards e.g. ISO 9241 or ISO 9126. During recent years the subjective perception of interactive systems is gaining more and more focus. There are many publications arguing that the user satisfaction dimension is equally important as other usability aspects (Tractinsky et al., 2000; Hassenzahl, 2004; Lavie & Tractinsky, 2004). A number of researchers consider it even more significant since there is some evidence that people's subjective attitudes can influence the perception of other usability components (Norman, 2002; 2004). These results may indicate that the role of the subjective assessment is more important in the usability evaluation than it was previously considered. Therefore, determining the real structure of preferences, understanding the way they are constructed or how they change over time should be a part of the main research interests of HCI scientists and practitioners. The need for such research was also advocated by Hornbæk's review paper (2006).

The users' preferences in the current study were investigated in the context of a simple 'point and click' method of controlling the application interface. Searching for various graphical objects and selecting them by a computer mouse click is a part of the human-computer interaction style called direct manipulation (Shneiderman, 1982; 1983). Despite enormous progress in the HCI field, this type of a man-machine dialogue is continuously among the most popular and extensively used in contemporary graphical interfaces. Some even argue (Whittaker et al., 2000) that too much focus is given to modern interaction styles whereas usefulness of these proposals is seriously limited (Hartson, 1998).

The research related to the graphical characteristics of a target object and their impact on human-computer interaction task efficiency can be divided into three general areas. The first trend focuses on the movement time needed to select a given object by means of a pointing device. The target is constantly visible for the user, so the recorded time expresses only the visually controlled motor activity. The process is described by the classical Fitts's law (Fitts, 1954; Fitts and Peterson, 1964).

A comprehensive review of numerous subsequent studies regarding Fitts's law was presented by MacKenzie (1991, 1992) and later by Plamondon and Alimi (1997). The second trend concerns studies in which the user performs the visual search in a graphical interface for a particular target among the group of distractors. These types of investigations were conducted under various conditions without the need of clicking the specified element. They stemmed from the observation that Fitts's law is not always sufficient in explaining the user behaviour since cognitive components may also influence the efficiency in the human-computer interaction (Card et al., 1983). It seems that one of the first papers in this area was the work of Backs and colleagues (1987). They asked the subjects to find a target object in vertical and horizontal menus and report an associated numerical value. The third field involves a combination of the first two groups. In this trend, the task completion required both the target search and selection. The rationale for such an approach is justified by the fact that the movement time in visually controlled activities and the time necessary for the visual search may not add up. A number of studies in this area can also be identified. One of the first researches was described in the paper of Deininger (1960). He studied the performance of keying telephone numbers using various arrangements of numerical keys. Many similar subsequent studies conducted before 1999 were in-depth reviewed by Kroemer (2001). A more detailed description and further references on these three trends are provided in the work of (Michalski et al., 2006).

The experimental task applied in the present work can be situated in the third trend described above and involves both a mouse pointer movement as well as a visual search component. The previous research results in this area were not consistent especially with respect to the location of the target object or group of objects on the screen. For instance, in the work of Campbell and Maglio (1999) it was shown that in 'search and click' tasks the shortest mean reaction times were observed for the stimuli located in the upper left corner of the screen. The worst results were recorded for targets positioned in the lower right corner. This finding was later supported by Schaik and Ling (2001). In their study on positioning the menu in web pages, the superiority of the left menu location over its right position was presented. However, later in a very similar study described by Pearson and Schaik (2003) the difference in reaction times between left and right menus for visual search tasks was not statistically meaningful. Moreover, McCarthy and colleagues (2003) in their eye tracking study provided some evidence that the left menu location is better in terms of the visual search task completion time, but only on the first visit to the www page. When the subject visited the web page again, there was no difference in mean acquisition times. The menu location effect was also insignificant in the study carried out by Kalbach and Bosenick (2004). The inconsistent results were also yielded when graphical toolbars were investigated. Michalski et al. (2006) showed that for horizontally oriented layouts there were no statistically significant differences in acquisition times between the left and right located arrangements. The outcome regarding the vertical panels was inconsistent though the differences in mean completion times were significant.

The contradictory results briefly described above show that the location effect probably strongly depends on the applied stimuli and the context of the research. In cases when the location seems to be unimportant, probably the users' preferences and subjective feelings play a more significant role in the usability evaluation of the given graphical interface. Although the efficiency and effectiveness of performing search and select tasks in various configurations of toolbars were also studied by Grobelny et al. (2005), Michalski & Grobelny (2006 and 2008), none of these researches included the subjective opinions. Therefore, in the present study, the selected geometrical features of vertical graphical structures were examined using both objective and subjective measures, and the focus is especially directed to increase the knowledge on the nature of the users' preferences towards simple graphical panels. More specifically, the following issues are addressed:

- Which of the geometric characteristics of the examined panels of icons are more important to the users before accomplishing the visual search and click tasks and after such an experience.

- Whether or not the users' preferences can change during the comparatively short period of time - directly after the interaction with simple search and point tasks. If so, what is the character of these changes.

- What is the relation between objective efficiency and effectiveness measures and subjective preferences while performing simple search and select tasks.

It is naturally hard to recommend one best approach to answer the research questions since every technique has its advantages and limitations. For that reason, a multi method approach might be a good solution to get a fuller picture of investigated issues. The objective measures related to the 'search and confirm' task performance (task completion times and errors made) were gathered by the custom made software and analysed by appropriate statistical procedures. For the purpose to assess the subjects' preferences towards experimental conditions, the Analytic Hierarchy Process method (Saaty, 1977; 1980) was used. The application of this approach, apart from providing relative weights, allowed additionally for the verification of the participants' preference consistencies. The preferences were measured twice which enabled the examination of their change dynamics. The relative weights obtained from the AHP were investigated by means of the conjoint analysis, which gave additional information on the aggregate-level relative importances for the examined factors. Moreover, the decision rules that are usually applied within this framework give the notion of the users' possible behaviour. The application of this method also extends the possibility of analysing the change of users' preferences both on an aggregate level as well as in the context of predicting the subjects' behaviour. It is quite obvious that people may differ from each other especially when it comes to expressing their preferences. Thus, to check whether there are some groups of participants that share the same opinions, the cluster analysis was employed. This approach was applied twice for the relative weights obtained prior and after the tasks performance. The detailed description of the carried out experiment and the obtained results along with the discussion and conclusions are provided in the remainder of the article.

# 1. Method

## 1.1. Participants

A total of 68 volunteers participated in the study. All reported having normal or corrected to normal visual acuity. All of the subjects attended courses organized by the Computer Science and Management Faculty at the Wroclaw University of Technology. There were fewer male participants (25 subjects, 37%) than females (43 subjects, 63%). All of the students were young within the age range of 21–25 years. The vast majority (56) of the volunteers worked with computers on a daily basis, whereas the rest (12) at least several times a week.

## 1.2. Apparatus

A custom computer program was used to carry out the experiments. The software was written in a MS Visual Basic™ 6.0 environment and the data was stored in a relational MS Access™ database. The research was carried out in teaching laboratories on uniform personal computers equipped with identical optical computer mice and 17" monitors of the CRT type. On all computer screens, the resolution was set at 1024 by 768 pixels and typical (default) computer mouse parameters were applied.

## 1.3. Independent variables

All of the examined structures comprised of 36 buttons with 26 Latin alphabet characters and ten Arabic numbers placed on these buttons. Bolded Times New Roman font types in three different sizes 12, 18 and 24 pts corresponding to the button dimension were used. The rationale for selecting letters as stimuli was to minimize the mental effort related with analysing the graphical structure of the target object. They are easily recognised and therefore the cognitive component can be attributed almost exclusively to the search process. These types of stimuli are also widely used in many visual search studies for similar reasons. However, to include the context of the human-computer interaction the letters were placed on standard Microsoft computer buttons arranged like toolbars in many Graphical

User Interfaces. The panels of graphical objects were differentiated according to the following three factors:

*Panel location on the screen.* The examined panels were situated in two different locations: the upper left and the upper right screen corners. They were moved away from the screen edges by 18 pixels to minimize the effect of a faster selection of items located at the screen borders (Farris et al., 2002, 2006; Jones et al., 2005). The applied gap corresponds to the typical height of the top title bar in many dialogue windows used in Microsoft® operating systems.

*Panel layout.* Four types of arrangements were analysed: squares with a side size of six elements (06_06); and three types of vertical rectangles: 18 rows and two columns (18_02), 12 rows and three columns (12_03), and nine rows and four columns (09_04).

*Graphical object size.* Three target sizes were used: small, medium, and large. The side button sizes amounted to 22, 30, and 38 pixels. Approximate visual angles of these items were equal to 0°41', 0°55', and 0°69' respectively. The standard square buttons that were employed in this study are widely utilized in many contemporary computer programs.

The examined graphical panels are illustrated in Figure 1. The picture presents 12 out of 24 experimental conditions. In the remaining 12, the only difference is that the respective arrangements are mirrorly placed in the right upper corner of the screen. The illustrations allow for assessing the proportion of the screen occupied by the examined structure.

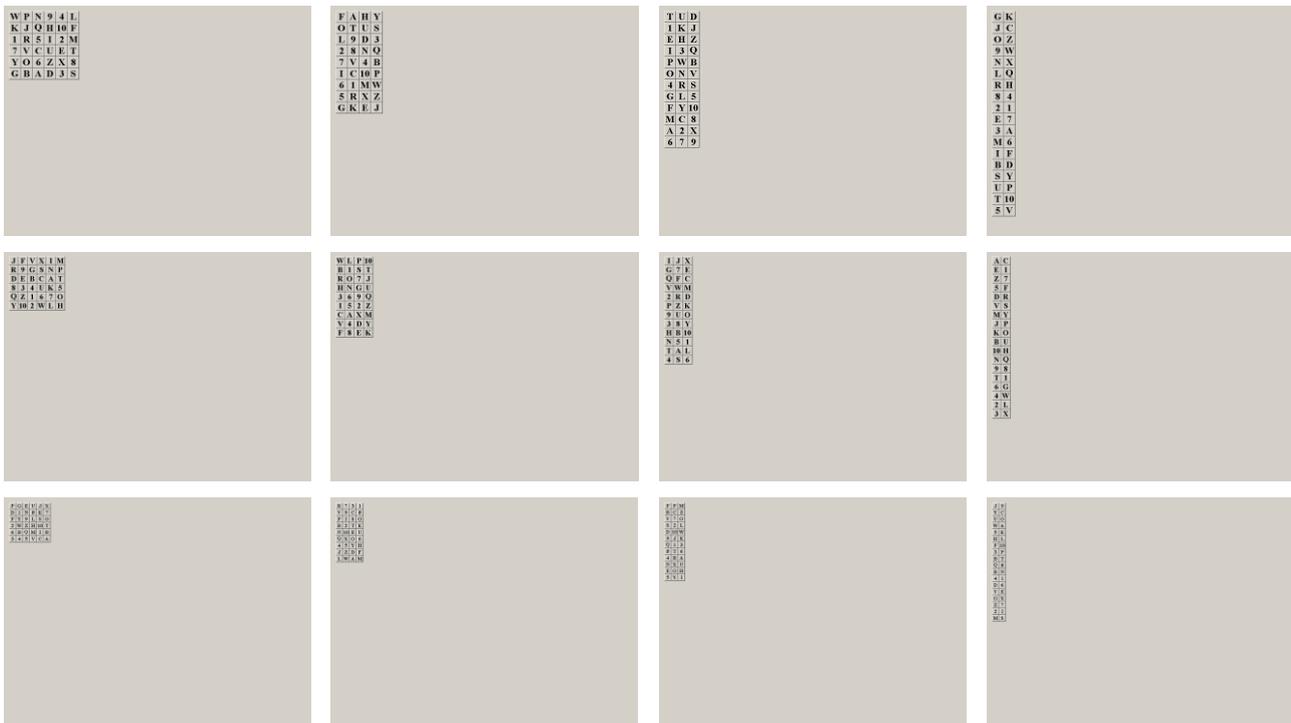

Figure 1. Twelve out of 24 experimental conditions used in this study. The remaining 12 looked exactly the same, but were placed in the right upper corner of the screen.

## 1.4. Dependent measures

The dependent variables being measured were twofold in nature. The first group regarded the task efficiency measures and included the acquisition time and the number of errors made. The time was computed from when the START button was pressed, to when the object was selected. The error occurred when the user selected a different than required graphical object.

The second set of recorded data was related to users' preferences expressed towards examined graphical structures. Generally, the methods of collecting the preferences may be divided into two main categories: objective and subjective. The objective ones include the examination of the human body's physical response to the stimulus (heart rate, skin conductivity, eye tracking study or most recently the imaging methods: PET, fMRI, MEG). They have many advantages, however, the subjective methods are easier to implement and are substantially cheaper and less troublesome. Among the subjective techniques, there are multiple ways of obtaining the subjects' preferences concerned with diverse objects or phenomena. The most common ones include direct ranking and pairwise comparisons. In this study the latter method was applied. The decision results from the findings published by Koczkodaj (1998), who showed the extreme superiority of the pairwise comparison techniques compared to the direct ranking with respect to the accuracy of the stimulus estimation. Although pairwise comparisons provide a better approximation of true preferences they have not been extensively used in the human-computer interaction field. The reason may lie in relatively more complex computations than in the direct rank approach. The other obstacle may be connected with the burden imposed on subjects, since the number of necessary comparisons grows rapidly with the number of analysed variants.

### 1.4.1. Priorities derivation

The responses given by the participants during the pairwise comparison procedure need to be processed to obtain the hierarchy of preferences. The most popular methods of calculating a priority vector from the numerical pairwise comparison matrix include the eigenvalue/eigenvector approach and the logarithmic least squares procedure (Dong et al., 2008). There is still ongoing discussion among the scientists which of the two approaches is better (e.g. Barzilai, 1997; Saaty & Hu, 1998). In this research, the Analytic Hierarchy Process (AHP) framework proposed by Thomas Saaty (1977, 1980) was applied to assess subjects' preferences. The AHP has been extensively used in various areas and the review of its applications can be found, for instance, in papers of Zahedi (1986) and Ho (2007).

The crucial idea of this multiple criteria decision tool lies in obtaining the hierarchy of the subjects' preferences from the symmetric and reciprocal pairwise comparison matrix by finding its principal eigenvector corresponding to the maximal eigenvalue of this matrix. More specifically, this square $n \times n$ matrix $\boldsymbol{P} = [p_{ij}]$ is constructed in such a way that for every $i, j = 1, ..., n$, $p_{ij} > 0$, $p_{ii} = 1$; $p_{ij} = (1/p_{ji})$, where $n$ is the number of alternatives, and $p_{ij}$ denotes the relative preference of an alternative $a_i$ over $a_j$ ($p_{ij} = a_i / a_j$). Thus, the priority vector of relative importances for individual alternatives $\boldsymbol{w} = [w_1, ..., w_n]^T$ is obtained from solving the following equation:

$$\boldsymbol{P} \cdot \boldsymbol{w} = \lambda_{max} \cdot \boldsymbol{w},$$

where the scalar $\lambda_{max}$ is the maximal eigenvalue of the matrix $\boldsymbol{P}$. The priority vectors computed individually for every subject were treated as a main dependent measure in this investigation. The higher the value of a relative weight, the bigger the preference for a given alternative. In the AHP approach the weights are normalized, that is why the dependent variable takes the values between zero and one, and the sum of all weights for a given participant equals one.

### 1.4.2. Consistency analysis

The significant advantage of the AHP framework is the possibility of estimating the consistency of responses for a particular individual. These consistency ratio values ($CR$) were calculated and analysed according to the formulas recommended in this method, namely: $CR = CI / RI$, where Consistency Index is computed according to the formula $CI = (\lambda_{max} - n) / (n - 1)$, and the Random consistency Index ($RI$) is dependent on the number of variants being compared (Forman, 1990). In this study the $RI$ amounted to 1.40.

Higher values of the $CR$ parameter correspond with the bigger inconsistencies in pairwise comparisons. Saaty (1980) considers a small level of inconsistency as a component necessary to achieve new knowledge. On the other hand, of course, the ratio cannot be too big and he proposed to set the upper acceptable $CR$ value at the level of 0.1.

## 1.5. Experimental design

Three independent variables resulted in twenty four different experimental conditions: (three object sizes) × (two panel locations) × (four layouts). The AHP approach relies on the pairwise comparisons, which the number ($c$) grows rapidly with the number of available options ($n$), namely $c = (n^2 - n) / 2$. The application of the within subjects design in this case would have resulted in as many as 276 comparisons. Therefore, to keep the amount of necessary comparisons reasonable a mixed model design (between and within subjects) was used to investigate all of the 24 sets of objects.

The decision as to what particular kind of experimental design to apply was generally based on the following two recommendations put forward by Saaty (1980), that is (1) to examine 'about seven elements' at the same time and (2) to group the variants according to the magnitude of their differences. Given the Fitts's law and the outcomes of the obtained by Michalski et al. (2006), the target object size factor was expected to differentiate the preferences the most. Hence, the object dimension factor was treated between subjects whereas the other two effects were examined within subjects. As a result of this, each of the three groups of participants were exposed to eight various experimental conditions.

## 1.6. Procedure

Participants were informed about a goal and a detailed range of the research. A block diagram illustrating the experimental procedure is shown in Figure 2.

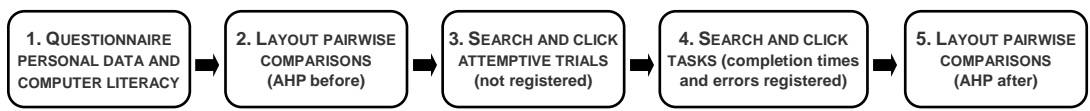

Figure 2. Block diagram of the experimental procedure

The study began by filling out a questionnaire regarding personal data and computer literacy. Next, subjects expressed their opinions on which of the further examined layouts would be better operated in terms of simple 'search and point' tasks. This evaluation stage was conducted by performing pairwise comparisons. The persons being examined specified their preferences pertaining to panels presented in a random order. The whole process was supported by the utilized software (Figure 3).

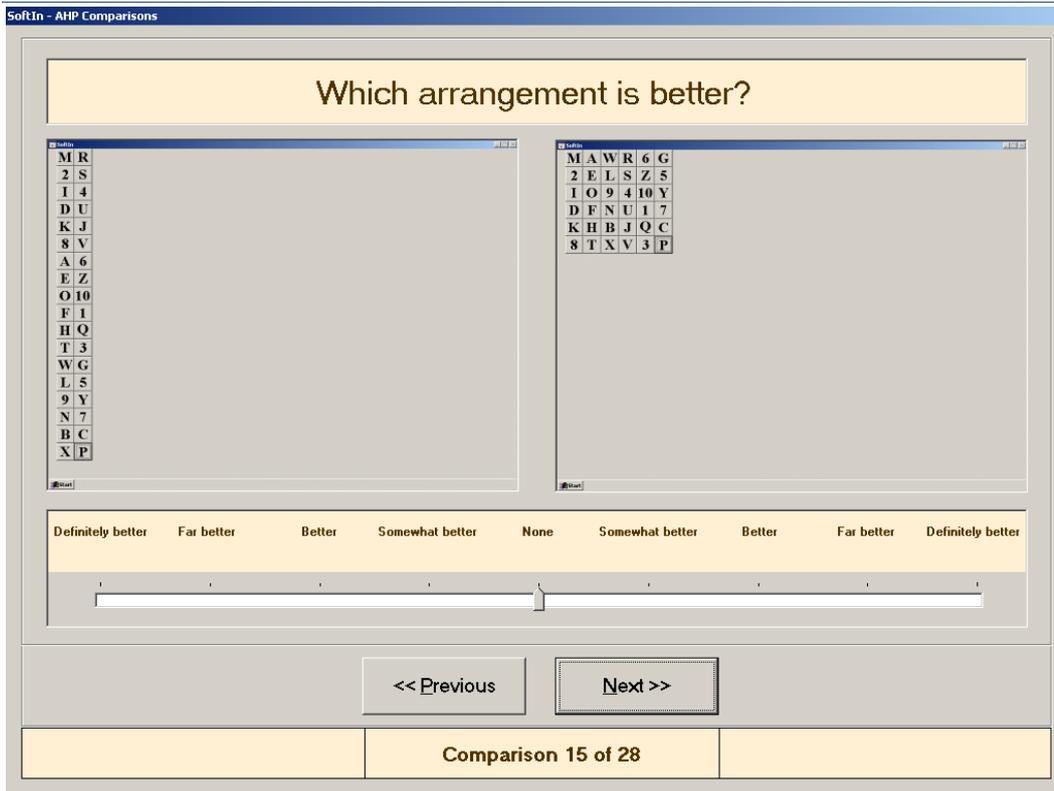

Figure 3. Exemplary pairwise comparison window in utilized software

Directly before the experiment, each subject executed attemptive trials which were not registered. The find and click task began by displaying a dialogue window with a START button and the target to be searched for. At this moment the layout was not visible to the user. After clicking the START button, the instruction window disappeared and one of the configurations was shown. The subject's task was to find and select by clicking a required item from among a group of randomly positioned distractors as fast as possible (Figure 4).

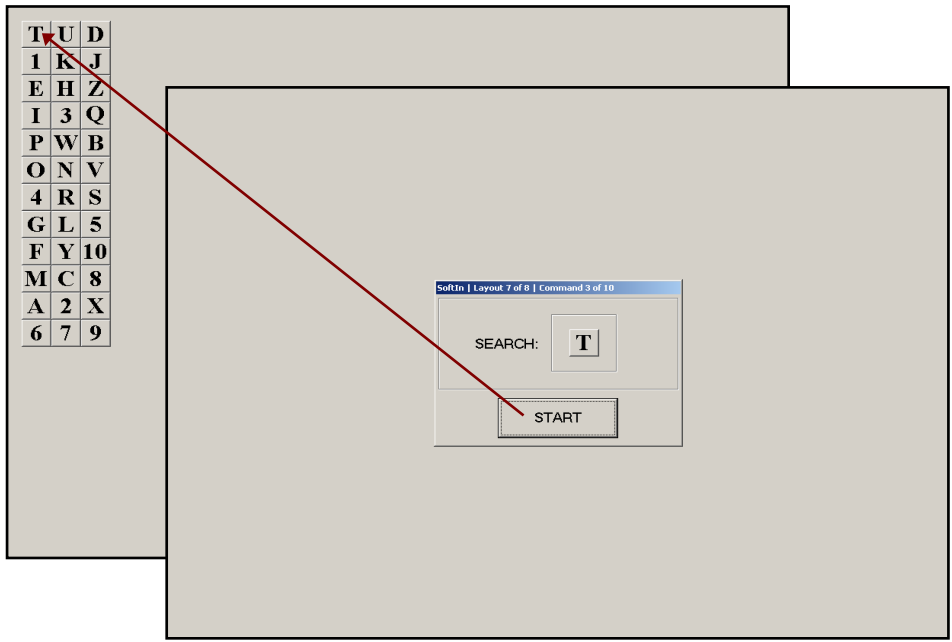

Figure 4. The example of the single search and click task

The instruction window appeared for each trial so every time the participant had to click the START first and then the searched item. The order of layout presentation was randomly set for every participant. The target letters were also chosen randomly as well as their location within the given panel. The letter once drawn for a particular toolbar did not appear again during one block of trials. Every subject performed 10 trials for each of the analysed conditions. The informative dialogue window including the average acquisition time and the number of incorrect attempts was shown every 10 trials. The average acquisition times included only properly selected elements. The subjects were not given immediate feedback about their mistakes to avoid losing their attention during executing the block of ten trials, diminish the possibility of excessive visual fatigue, and prevent additional mouse clicking.

All 'search and click' tasks were executed by means of a standard computer mouse. The distance between the user and the computer monitor was set approximately at 50 cm. After conducting the efficiency examination subjects repeated the same subjective evaluation procedure which had been administered before performing the 'find and click' trials.

## 2. Results

### 2.1. Objective measures

The following subsections deal with the objective measures of the users' performance during the search and click trials. The descriptive statistics along with the analysis of variance was applied to examine the registered acquisition times. The errors made during performing the experimental tasks were characterized and analysed by means of the appropriate statistical test.

### 2.1.1. Descriptive statistics

General descriptive characteristics of all acquisition times including central tendency and variability measures, along with shape characteristics are put together in Table 1. From these data it can be easily noted that the individual central tendency parameters differ markedly one from another. The mean value is as much as 30% bigger than the median value. For the normal distribution these two parameters have similar values. The considerable differences between the obtained descriptive characteristics and the parameter values characteristic of the Gaussian distribution are also visible in the calculated skewness (3.5) and kurtosis (21). The positive value of the skewness means that more of the variate values are located on the left hand side of the distribution than on its right part. In turn, the large kurtosis value estimated from the sample suggests that the probability density distribution in this case is significantly less dispersed than the normal distribution.

Table 1. General descriptive characteristics of all acquisition times

| Parameter | Value |
|---|---|
| Valid cases | 5361 |
| Means | |
|     Arithmetic | 2194 [ms] |
|     Geometric | 1808 [ms] |
|     Harmonic | 1559 [ms] |
| Median | 1683 [ms] |
| Minimum | 491 [ms] |
| Maximum | 23 764 [ms] |
| Variance | 2 977 859 [ms]$^2$ |
| Standard Deviation | 1726 [ms] |
| Mean Standard Error | 23.6 [ms] |
| Skewness | 3.5 |
| Kurtosis | 21 |

The basic statistical parameters calculated for every experimental condition are given in Table 2. These results present characteristics similar to those obtained for the whole variate, e.g. for every examined layout the median is much lower than the mean value.

Table 2. Descriptive statistics of acquisition times for all experimental conditions

| Location | Layout | Small | | | | | Medium | | | | | Large | | | |
|---|---|---|---|---|---|---|---|---|---|---|---|---|---|---|---|
| | | N[*] | Mean (MSE[**]) | Median | Min | Max | N | Mean (MSE) | Median | Min | Max | N | Mean (MSE) | Median | Min | Max |
| Left | 06_06 | 219 | 2332 (118) | 1832 | 751 | 14 110 | 229 | 2048 (118) | 1653 | 621 | 16 444 | 228 | 1744 (75) | 1387 | 611 | 7 971 |
| | 09_04 | 218 | 2499 (132) | 1858 | 701 | 16 634 | 225 | 2139 (118) | 1523 | 671 | 14 361 | 225 | 1768 (70) | 1492 | 661 | 8 011 |
| | 12_03 | 216 | 2377 (104) | 1958 | 771 | 9 614 | 226 | 2360 (124) | 1682 | 751 | 11 036 | 227 | 1904 (81) | 1542 | 671 | 8 131 |
| | 18_02 | 217 | 2540 (116) | 1973 | 791 | 12 498 | 225 | 2353 (120) | 1742 | 591 | 12 708 | 223 | 2164 (112) | 1633 | 621 | 10 185 |
| Right | 06_06 | 217 | 2385 (117) | 1833 | 640 | 13 690 | 224 | 1917 (81) | 1583 | 671 | 8 682 | 224 | 1924 (96) | 1423 | 491 | 9 533 |
| | 09_04 | 217 | 2390 (129) | 1792 | 721 | 14 331 | 225 | 2007 (92) | 1462 | 691 | 8 051 | 227 | 1802 (101) | 1331 | 621 | 11 126 |
| | 12_03 | 218 | 2483 (126) | 1978 | 701 | 13 749 | 227 | 2221 (142) | 1722 | 630 | 23 043 | 230 | 1962 (109) | 1572 | 601 | 20 119 |
| | 18_02 | 219 | 2847 (168) | 2083 | 721 | 23 764 | 227 | 2476 (137) | 1813 | 691 | 15 452 | 228 | 2114 (109) | 1653 | 581 | 13 269 |

[*] N – number of valid cases
[**] MSE – Mean Standard Error

The shortest mean acquisition times (1744ms) were registered for the square layout (06_06) comprising the largest target items, located in the left upper corner of the computer screen. The worst average results were obtained for the most vertical layout (18_02) positioned in the right upper corner of the monitor, consisting of the smallest objects (2847ms). The relative difference between mean 'search and click' times of these two layouts amounted to 60%.

## 2.1.2. Analysis of variance

The descriptive statistics presented and analysed in the previous section of this paper clearly show that the acquisition time variate empirical distribution in this study is rather unlikely to come from the Gaussian distribution. Therefore, instead of the standard analysis of variance, an ANOVA available within the Generalized Linear Models (GZLM; Nelder and Wedderburn, 1972) was conducted under the assumption that the dependent variable has the inverse Gaussian (IG) distribution. This approach seems to be justified in light of the work of Michalski (2005), where it was shown that the hypothesis about the IG character of the acquisition time empirical distribution for the types of graphical structures comparable with the examined in this study cannot be rejected. Thus, a three factorial analysis of variance based on the GZLM was employed for assessing the effects of the panel location on the screen, its layout, and target sizes. The results of this analysis are summarized in Table 3 and graphical illustrations of mean acquisition times obtained for the significant factors are presented in Figures 5 and 6.

Table 3. Analysis of variance results related to acquisition times

| Factor | df | Wald statistics | p |
|---|---|---|---|
| Location | 1 | 0.34 | 0.56 |
| Target size | 2 | 149 | < 0.00001[*] |
| Layout | 3 | 52 | < 0.00001[*] |
| Location × Target size | 2 | 3.5 | 0.18 |
| Location × Layout | 3 | 2.6 | 0.46 |
| Target size × Layout | 6 | 6.9 | 0.33 |
| Location × Target size × Layout | 6 | 6.2 | 0.41 |

[*] $p < 0.00001$

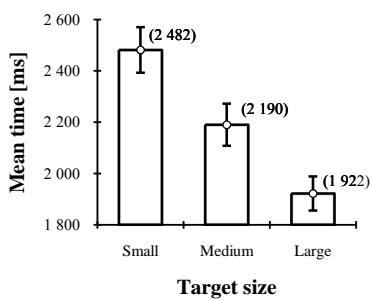

Figure 5. Mean acquisition times depending on object size ($W = 149$, $p < 0.00001$). Whiskers denote mean standard errors.

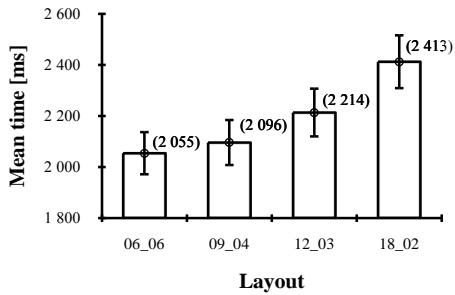

Figure 6. Mean acquisition times depending on panel layout ($W = 52$, $p < 0.00001$). Whiskers denote mean standard errors.

Means depending on the panel location did not differ considerably (*Wald statistics* ($W$) = 0.34, $p = 0.56$), whereas the effects of the object size ($W = 149$, $p < 0.00001$) and panel layout ($W = 52$, $p < 0.00001$) were statistically significant. None of the possible interactions was meaningful. Figure 5 shows a tendency of decreasing the task accomplishment times along with the increase of the target size. Also quite a clear trend can be observed in Figure 6 where longer mean completion times were registered for more vertical panel layouts.

Effects that significantly affected the acquisition times, namely the object size and the panel layout, were additionally post-hoc explored by means of GLZM one way ANOVAs. The results are presented in Tables 4 and 5 respectively.

Table 4. Series of GLZM one way ANOVAs for the target size factor (*Wald* statistics, $p$ values in brackets)

|  | **Small** | **Medium** | **Large** |
|---|---|---|---|
| **Small** | × | 34 (0.00001) | 150 (0.00001) |
| **Medium** |  | × | 39 (0.00001) |
| **Large** |  |  | × |

Table 5. Series of GLZM one way ANOVAs for the layout factor (*Wald* statistics, $p$ values in brackets)

|  | **06_06** | **09_04** | **12_03** | **18_02** |
|---|---|---|---|---|
| **06_06** | × | 0.70 (0.40) | 9.7 (0.0019) | 42 (0.00001) |
| **09_04** |  | × | 5.1 (0.024) | 31.5 (0.00001) |
| **12_03** |  |  | × | 11.8 (0.0006) |
| **18_02** |  |  |  | × |

In the case of the target size, statistically meaningful ($p < 0.05$) differences were detected between all pairs of factor levels. When the configuration of the panel is concerned the only insignificant difference was identified between the square layout and the most compact vertical panel with four columns and nine rows ($W = 0.70$, $p = 0.40$). The rest pairwise comparisons were statistically important ($p < 0.05$).

### 2.1.3. Errors analysis

The subjects made 79 errors altogether, which accounts for 1.5% of all trials. A nonparametric Chi-square test was employed to verify the significance of differences in the number of wrong selections for the independent variables. The effects of the panel configuration, location, and target size did not influence the number of errors made (the significance level = 0.05). There were also no statistically meaningful differences in the number of mistakes between male and female participants.

## 2.2. Preference analysis

The subjective results obtained by means of the Analytic Hierarchy Process are presented in this section. They include the consistency ratio analysis for subjects taking part in the examination along with the detailed analysis of the calculated preferences.

### 2.2.1. Concordance of preferences

In the conducted experiment the consistency ratio values ranged from 0.009 up to even 0 .401 with the overall mean of 0.132. A number of descriptive statistical characteristics of the consistency ratios computed in this investigation for all experimental conditions are put together in Table 6.

Table 6. Descriptive statistics of consistency ratio values for all experimental conditions

| Gender | Target size | Time | Mean CR | MSE | Min | Max | N | No. of subjects with CR < 0.1 |
|--------|-------------|------|---------|-----|-----|-----|---|-------------------------------|
| Women | Small | Before | 0.133 | 0.027 | 0.031 | 0 .364 | 14 | 7 |
| | | After | 0.163 | 0.023 | 0.051 | 0 .343 | 14 | 4 |
| | Medium | Before | 0.143 | 0.019 | 0.047 | 0 .292 | 14 | 4 |
| | | After | 0.136 | 0.027 | 0.040 | 0 .401 | 14 | 6 |
| | Large | Before | 0.149 | 0.018 | 0.067 | 0 .296 | 15 | 5 |
| | | After | 0.164 | 0.022 | 0.056 | 0 .325 | 15 | 4 |
| Men | Small | Before | 0.083 | 0.020 | 0.009 | 0 .161 | 8 | 5 |
| | | After | 0.100 | 0.019 | 0.029 | 0 .215 | 8 | 6 |
| | Medium | Before | 0.124 | 0.022 | 0.062 | 0 .284 | 9 | 3 |
| | | After | 0.110 | 0.015 | 0.054 | 0 .182 | 9 | 5 |
| | Large | Before | 0.123 | 0.028 | 0.018 | 0 .244 | 8 | 4 |
| | | After | 0.157 | 0.036 | 0.045 | 0 .319 | 8 | 3 |

[*] N      – Number of valid cases
[**] MSE  – Mean Standard Error

It might be interesting to find out whether the effects of gender, target size, and the moment of making the pairwise comparisons (before or after the performance tasks) had an impact on the reliability of the obtained subjective results. For this purpose the standard analysis of variance was employed. Its results are summarized in Table 7.

Table 7. Consistency ratio analysis of variance results

| Factor | df | F | p |
|--------|----|----|----|
| Gender | 1 | 4.8 | 0.03[*] |
| Target size | 2 | 1.3 | 0.27 |
| Moment of comparisons (MOC) | 1 | 0.78 | 0.38 |
| Gender × Target size | 2 | 0.72 | 0.49 |
| Gender × Moment of comparisons | 1 | 0.000034 | 0.995 |
| Target size × Moment of comparisons | 2 | 0.66 | 0.52 |
| Gender × Target size× MOC | 2 | 0.12 | 0.89 |

[*] p < 0.05

Only the differences between mean *CR*s in relation to the gender effect occurred to be statistically meaningful ($F = 4.8$, $p = 0.03$). The women were generally less coherent ($CR = 0.15$) in expressing their preferences towards examined graphical structures than men ($CR = 0.12$). However, in this case the ANOVA was not fully balanced because there were more women than men taking part in the examination (compare the eighth column in Table 6). The other two analysed factors along with all the interactions happened to be insignificant.

The last column from the Table 6 contains the number of participants that have met the criterion of $CR < 0.1$ in the current study. In order to formally verify if there were any significant differences between the number of subjects excluded from the further preference analysis in relation to the factors of a gender, target size, and the moment of retrieving the relative weights (before or after the performance tasks), the Chi-square statistical test was used. The results can be found in Table 8.

Table 8. The Chi-square test results of the number of subjects excluded from the AHP weight analysis in relation to gender, target size and moment of comparisons factors

| Factor | df | Chi-square | p |
|---|---|---|---|
| Gender | 1 | 3.8 | 0.05[*] |
| Target size | 2 | 2.3 | 0.32 |
| Moment of comparisons (MOC) | 1 | 0 | 1 |

[*] p <= 0.05

The conducted Chi-square tests confirm the analysis of variance outcomes presented in Table 7. A significantly bigger proportion of women (65% exclusions) than men (48%) did not obtain consistency ratio values below the recommended by Saaty's threshold.

## 2.2.2. Subjective weights ANOVA

The results pertaining to the users' preferences towards examined structures both before and after the efficiency examination are presented in this section. The relative likings are expressed as average values of the obtained AHP weights. The demonstrated findings encompass only the results with consistency ratios lower than 0.1 and the preferences meeting this condition are summarized in Table 9.

Table 9. Mean AHP weights (preferences) for all experimental conditions (mean standard error in brackets)

| Target location | Layout | Object size | | | | | |
| | | Small | | Medium | | Large | |
| | | Before | After | Before | After | Before | After |
|---|---|---|---|---|---|---|---|
| Left | 06_06 | 0.128 (0.019) | [*]0.169 (0.032) | 0.107 (0.024) | [**]0.140 (0.023) | [*]0.105 (0.021) | 0.144 (0.031) |
| | 09_04 | [**]0.145 (0.021) | [**]0.174 (0.023) | [*]0.103 (0.012) | 0.123 (0.016) | 0.119 (0.021) | [**]0.145 (0.028) |
| | 12_03 | 0.142 (0.015) | 0.139 (0.020) | [**]0.140 (0.017) | 0.115 (0.013) | 0.149 (0.014) | 0.120 (0.017) |
| | 18_02 | [*]0.120 (0.022) | [*]0.065 (0.007) | 0.133 (0.032) | [*]0.104 (0.019) | [**]0.217 (0.037) | [*]0.105 (0.021) |
| Right | 06_06 | 0.118 (0.022) | [~~]0.133 (0.024) | [~]0.100 (0.017) | 0.134 (0.025) | [~]0.088 (0.016) | [~~]0.137 (0.032) |
| | 09_04 | [~]0.110 (0.011) | 0.124 (0.013) | 0.136 (0.024) | [~~]0.135 (0.014) | 0.095 (0.013) | 0.115 (0.014) |
| | 12_03 | [~~]0.125 (0.016) | 0.121 (0.016) | [~~]0.163 (0.024) | 0.132 (0.015) | 0.089 (0.013) | 0.124 (0.015) |
| | 18_02 | 0.113 (0.024) | [~]0.074 (0.019) | 0.118 (0.020) | [~]0.117 (0.022) | [~~]0.139 (0.019) | [~]0.112 (0.030) |

[*]    The lowest mean weight value among layouts located on the left
[**]   The highest mean weight value among layouts located on the left
[~]    The lowest mean weight value among layouts located on the right
[~~]   The highest mean weight value among layouts located on the right

Although it is difficult to observe an obvious relationship among the calculated preference weights presented in Table 9, yet a certain tendency may be noticed. When the subjects rated the layouts before the experimental task, the bigger preferences were generally given to the more vertically oriented layouts. The lowest rates were given to the more compact structures. There was, however, one evident exception from this rule – small targets positioned on the left. This may probably be attributed to some users' habits since usually this is the most common location of many menus and toolbars in contemporary graphical interfaces. The preferences seem to change decidedly for the pairwise comparisons given after the 'search and click' tasks. In this case the highest preferences were assigned to the more compact structures, whereas the lowest values were registered for the more vertical layouts.

To obtain a more clear view on the aforementioned trend, the mean weight scores were computed across all target sizes used in this study in relation to the layout variable, separately for the results obtained before and after the experimental tasks. The graphical representation of these calculations is provided in Figure 7.

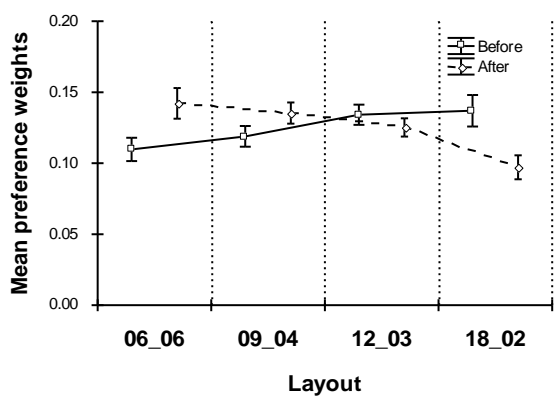

Figure 7. Graphical presentation of the mean preference weights depending on panel layout obtained before and after performance tasks (whiskers denote mean standard errors)

A similar approach was employed to the layout location variate, and the outcome is illustrated in Figure 8. Also in this case, the moment of carrying out the comparisons caused the shift in the registered mean weights. The Figure 8 shows that after the 'find and click' trials, there was a decrease in mean preferences for panels located on the left hand side of the screen, while the layouts positioned on the right were better rated after than before the experimental tasks.

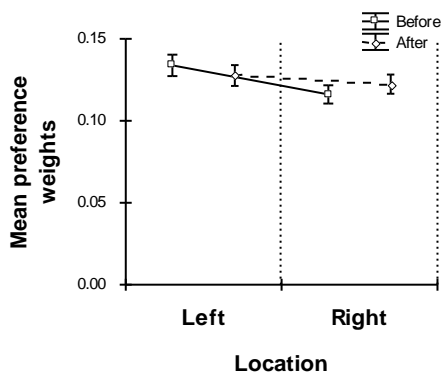

Figure 8. Graphical presentation of the mean preference weights depending on panel location obtained before and after performance tasks (whiskers denote mean standard errors)

In general, the repeated measures ANOVA would have been most suitable to formally verify if the presented above outcomes have any meaning. However, in this research, according to the Saaty's recommendation, only the weights obtained for participants with the consistency ratio lower than 0.1 were taken into account. The $CR$ values were obtained for each subject twice (prior and after the search and click trials) and it occurred that they were below that threshold in both cases for 16 out of 68 persons taking part in the experiment (23.5%). Thus, the repeated measures ANOVA could be conducted only on those 16 subjects whereas the consistent data ($CR < 0.1$) were gathered for 28 participants before the task performance and for (partially) different 28 subjects after the exposure (compare the two last columns in Table 6). Therefore, not to neglect the considerable amount of valid preference weights (more than 40%), the standard factorial instead of repeated measures ANOVA was employed. The two way analysis of variance including the effects of the layout and target location was used separately for the preferences before and after the task performance and the results are presented in Table 10.

Table 10. ANOVA analysis of preferences before and after the experimental trials

| Factor | df | Before | | After | |
|---|---|---|---|---|---|
| | | F | p | F | p |
| Layout | 1 | 2.3 | 0.082[*] | 5.5 | 0.0012[***] |
| Location | 3 | 4.3 | 0.039[**] | 0.38 | 0.54 |
| Layout × Location | 3 | 0.26 | 0.85 | 0.66 | 0.58 |

[*] p < 0.1
[**] p < 0.05
[***] p < 0.002

The panel layout variable was statistically meaningful before ($F = 2.3$, $p = 0.082$) but only at the significance level $a = 0.1$, and after performing the tasks at the level of 0.002 ($F = 5.5$, $p = 0.0012$). The location factor was significant ($a = 0.05$) solely before the efficiency examination ($F = 4.3$, $p = 0.039$), and the superiority of the left positioned layout disappeared after the 'search and click' tasks were performed. All of the interactions were irrelevant.

## 2.2.3. Conjoint analysis

The foundations of the conjoint analysis were elaborated by Luce & Tukey (1964) and Krantz & Tversky (1971) and used later successfully in various fields for analysing people's preferences (Green and Srinivasan, 1978, 1990; Green et al., 2001). The general idea of this approach is first to obtain the overall preferences towards examined variants and then to find partial contributions of their attributes. The advantages of applying this framework include the calculation of relative importances for attributes and so called part-worths for all attribute levels.

In the present study, the dummy variable regressions were applied within the conjoint analysis framework. The AHP relative weights obtained for every participant were used as a measure of an aggregate response for experimental conditions. The part-worths along with attribute relative importances were computed and the results are demonstrated in Table 11.

Table 11. Aggregate-level relative importances and part-worth estimates.

| Variables | Relative importance | | Part-worth estimates | |
|---|---|---|---|---|
| | Before | After | Before | After |
| **Location** | 39.5% | 12% | | |
| Left | | | 0.00889 | 0.00262 |
| Right | | | -0.00889 | -0.00262 |
| **Layout** | 60.5% | 88% | | |
| 06_06 | | | -0.01521 | 0.01721 |
| 09_04 | | | -0.00602 | 0.01038 |
| 12_03 | | | 0.00920 | -0.00024 |
| 18_02 | | | 0.01203 | -0.02784 |

The obtained aggregate-level relative importances show a considerable change of preferences towards tested panels. Prior to the testing phase, the relative importance of the layout factor was modestly bigger than the value for the location effect. The values amounted to about 60% and 40% respectively. After the search and click tasks, there was a substantial increase of the relative importance for the layout variable, and the proportions changed to 90% and 10%.

The $R$-squared values calculated for every subject's regression were used as a goodness-of-fit criterion. The mean value of this parameter equalled to 85.1% prior the search and point tasks and 86% after. The average for all of the regressions of the $F$ statistics amounted to 18.5 before and 32.3 after, with the average significance levels $p_{before} = 0.146$ and $p_{after} = 0.137$.

The conducted conjoint analysis is usually followed by applying some of the existing models that allow predicting the user's behaviour. Among the most popular decision rules are the first choice model (FCM), Bradley, Terry, and Luce (BTL) probability choice model, and logit probability model (LPM). The first choice model is based on the proportion of subjects that rated the given variant the highest in relation to the total number of subjects. In the Bradley-Terry-Luce model, the choice probabilities for a given subject are obtained by dividing the utility of the profile by the sum of all profiles' utilities. Next, the probabilities are averaged across all subjects. In this study, the geometric instead of arithmetic means were employed due to the recommendation provided in the AHP approach. In the logit model the choice probabilities are estimated in a similar way as in the BTL procedure but before the division the numerator is calculated by raising the Euler's constant to the power of the appropriate utility. The denominator is a sum of the Euler's constants raised to power of their respective utilities. Using the regression approach in the conducted conjoint analysis it is possible to obtain negative predicted preference weights which, of course, doesn't have any interpretation in the AHP framework. Therefore, persons for whom such values were found were not taken into account both in the BTL and LPM models. The Table 12 shows the results of these three simulation models and the graphical representation is given in Figures 9–11.

Table 12. Different choice model simulation results.

| No. | Location | Layout | FCM[*] | | BTL[**] | | LPM[***] | |
|-----|----------|--------|--------|-------|---------|-------|----------|---------|
| | | | Before | After | Before | After | Before | After |
| 1. | Left | 06_06 | 16.1% | 19.4% | 0.095 | 0.117 | 0.1240 | 0.12699 |
| 2. | Left | 09_04 | 9.7% | 16.1% | 0.112 | 0.120 | 0.1245 | 0.12628 |
| 3. | Left | 12_03 | 9.7% | 12.9% | 0.131 | 0.119 | 0.1265 | 0.12522 |
| 4. | Left | 18_02 | 25.8% | 0.0% | 0.115 | 0.084 | 0.1266 | 0.12183 |
| 5. | Right | 06_06 | 6.5% | 22.6% | 0.088 | 0.115 | 0.1232 | 0.12684 |
| 6. | Right | 09_04 | 6.5% | 3.2% | 0.104 | 0.128 | 0.1237 | 0.12598 |
| 7. | Right | 12_03 | 9.7% | 6.5% | 0.121 | 0.118 | 0.1257 | 0.12503 |
| 8. | Right | 18_02 | 16.1% | 22.6% | 0.108 | 0.069 | 0.1258 | 0.12182 |

[*]   First Choice Model
[**]  Bradley, Terry, and Luce Model
[***] Logit Probability Model

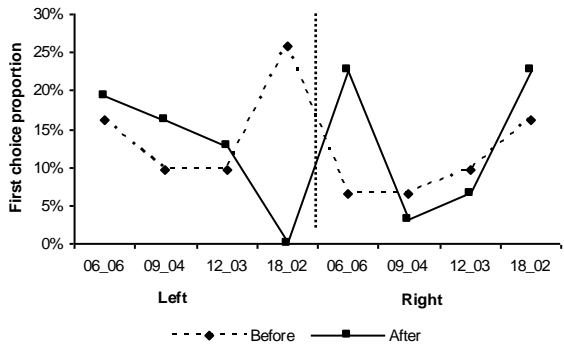

Figure 9. First Choice Model results.

In the first choice model, the results show different user behaviour for the toolbar positioned in the left and right upper corners of the screen. Initially, among the left located panels the biggest share of the first choices was attributed to the most vertical configuration (18_02). This was radically reversed after the experimental trials when none of the examinees selected this panel. Moreover, a gradual decrease in the first choice proportions along with the increase of the panel verticality could be observed for the remaining structures.

On the right hand side the first choice preferences before the efficiency examination were the bigger, the more vertical the toolbar was. After the search and click tasks the greatest change occurred for the most compact, square configuration. In addition, unlike in the left positioned structures, there was an unexpected increase in first choices of the most vertical panels.

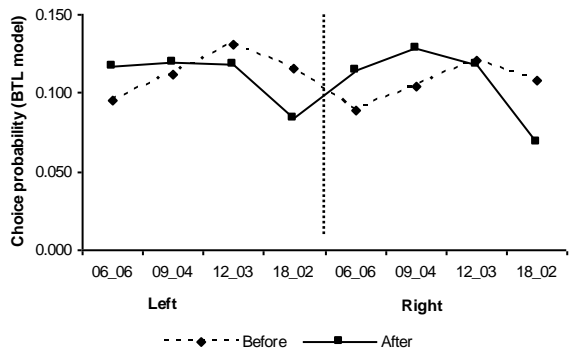

Figure 10. Bradley, Terry, Luce Model results.

When considering the values obtained by means of the BTL model one may notice that prior to the efficiency examination, the smallest probabilities were computed for the square structures, whereas the panels with the 12 rows and three columns located both in the left and right upper corners of the screen obtained the highest values. After the search and click tasks, the biggest increase in probabilities was observed for square panels, while the considerable drop was noticed for the most vertical configurations (18_02).

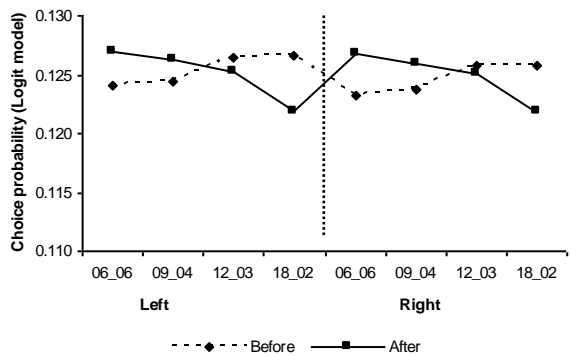

Figure 11. Logit Probability Model results.

The most regular and clear suggestions were provided by the LPM model. Before the tasks took place the choice probability was bigger for the more vertical configurations both in the left and right upper corners. After the search and click tasks the trend was apparently reversed suggesting bigger probabilities for the more compact structures.

Generally, the BTL and LPM choice models applied to the preference weights are quite similar. They favour the most vertical panels (12_03 and 18_02) situated both in the left and right upper corners of the screen prior to accomplishing the search and click tasks. In both approaches also a significant decrease in probabilities of choosing the toolbar with 18 rows and two columns was observed after testing the graphical structures.

### 2.2.4. Cluster analysis

To further explore the nature of the users' preferences towards examined toolbars, the k-means cluster analyses were carried out for the data gathered prior and after the search and click trials. The graphical illustration of the findings is presented in Figure 12 and 13 respectively.

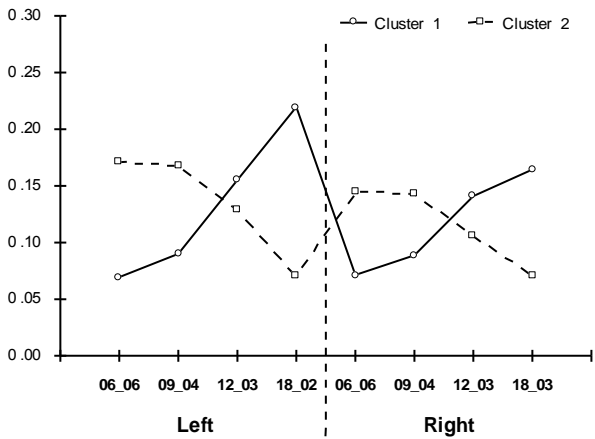

Figure 12. Plot of Mean weights for each cluster before search and point tasks

The cluster analysis of relative weights obtained prior to the pointing tasks suggests the existence of two groups of participants with distinct preference structures. The first cluster consists of 17 persons who are strong supporters of the verticality of the tested panels situated both in the left and right upper corner of the screen. The more vertical the panel is, the bigger the mean weight value is assigned to it. In the second group (14 subjects), generally, the bigger verticality of the toolbar corresponded with the smaller preference weights, however the difference in average values between the most compact configurations (square versus 09_04) is tiny.

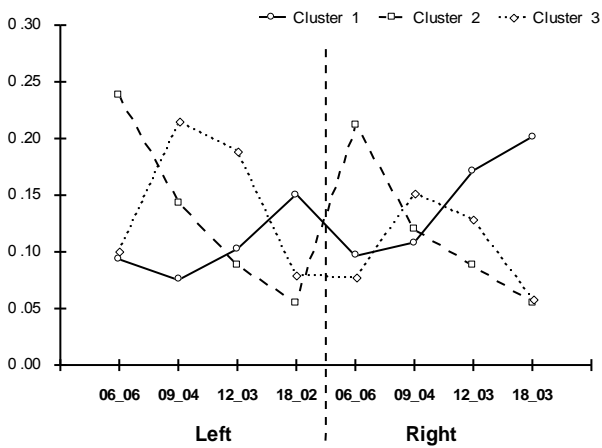

Figure 13. Plot of Mean weights for each cluster after search and point tasks

The clustering conducted after the search and click trials revealed three various groups of users. The first cluster (10 participants) is similar to the cluster 1 obtained before the panel testing yet, in this case, the mean weights for toolbars on the left hand side are generally lower than their counterparts prior to the pointing tasks. Surprisingly, the panels on the right were higher rated compared to their equivalents evaluated before the tasks were performed. The second cluster consists of 11 compact configuration admirers and is comparable with the second cluster from the previous analysis. However, this time, there is a much bigger difference between the mean preference weights towards the square structures and configurations with nine rows and four columns. In addition, the relation seems to be almost linear. The last cluster includes 10 people who apparently dislike square solutions but among the other, vertical configurations they prefer those more compact.

## 3.    Discussion

The present research was designed to provide some insight into nature of the preferences towards graphical configurations similar to widely used toolbars in conjunction with the simple direct manipulation tasks. Both subjective and objective measures were involved in this research. The gathered data were next thoroughly analysed by means of multiple methods and the discussion of the results is presented in the following subsections.

### 3.1.   Objective measures

The outcomes related to the target size factor that were obtained in this research confirm the selection time results reported, for example, in the studies of Sears and colleagues (1993) or Näsänen et al. (2001). They clearly show that the bigger the target stimuli are, the shorter acquisition times are obtained. However, it should be noted that the findings from the work of Grobelny et al. (2005) together with the model presented by Michalski et al. (2006) suggest that there may exist an optimal size of the graphical objects arranged in panels. Additionally, there is evidence that under some circumstances, the item size effect may be statistically irrelevant (e.g. Sears and Zha, 2003).

As for the layout variable, the efficiency results showed the superiority of more compact arrangements (e.g., squares) over the more dispersed layouts. This finding is fully in agreement with the work of Michalski et al. (2006) where similar panels were tested.

The acquisition times registered for the two panel locations did not differ significantly which is generally consistent with the findings reported by Michalski et al. (2006). However, they examined both horizontal and vertical configurations and the more detailed analysis shown meaningful (p < 0.05) discrepancies in two vertical cases: for the nine rows and four columns panels as well as for 12 rows and three columns ones. An analogous analysis was conducted using the data received in the present study. The series of one way GLZM analyses of variance showed no statistically significant ($a = 0.1$) differences in mean acquisition times, so the outcome is in contradiction with the findings provided by Michalski et al. (2006). The discrepancy may be due to the different experimental setup employed in these studies. Namely, in the present research users were presented only the vertical configurations plus the square one and the target size didn't vary within a group whereas in Michalski et al. (2006) investigation, the participants were exposed for much greater variations in experimental conditions. Maybe the variety of factors explored within subjects was continuously forcing the users to adopt different search strategies, which could have an impact on the efficiency performance. It is also plausible that the inclusion of the subjective evaluation into the procedure could have influenced the performance of the users by distracting their attention from the experimental tasks.

The finding that the target location is irrelevant when the simple 'search and click' task times are concerned is also contradictory with the works of Campbell and Maglio (1999) as well as Schaik and Ling (2001). On the other hand, the obtained in this investigation results regarding the location factor are in concordance with the findings presented by Pearson and Schaik (2003), Kalbach and Bosenick (2004) and partially by McCarthy and colleagues (2003). It should be noted, however, that experimental setups in those researches differed very much from the present study. In most of the cited papers the focus was directed to the web pages so they involved text hyperlinks as target objects. The selected context and a specific type of stimulus probably could have had a considerable impact on the results.

One should also be aware that during the experiment the instructive window, with the object to be searched for, was always displayed at the centre of the screen so the distances from the START button varied both across the layouts and within an individual configuration. However, the full control of the distance along with maintaining some degree of similarity between the experimental setup and the real-life would be quite difficult to achieve. Either the target would have to appear always in the same locations or the start button would have to change continuously its position on the screen which could be a little bit awkward for the user. Furthermore the main focus of the presented research was given to the preference examination, therefore the factors influencing the search and click times were of less importance than the preference related issues.

In general, the distance to the target could be of great importance but mostly in the case of involving visually controlled motor activities where the target is constantly visible to the participant. In the current study the task required both the target search and object selection so the task completion times depended both on the mouse movement time and the time necessary for the visual search. Usually the cognitive components are modelled by means of the Hick-Hyman's law (Hick, 1952; Hyman, 1953) and the visually controlled motor activities by the Fitts's law (Fitts, 1954; Fitts and Peterson, 1964). The Hick–Hyman's equation relates the reaction time (*RT*) to the set of *n* stimuli and takes the following form $RT = a + b \log_2(n)$, where *a* and *b* are empirically determined constants. The Fitts's law, in turn, says that the movement time (*MT*) to one constantly visible target is linearly proportional to the so called index of difficulty (*ID*): $MT = a + b \cdot ID$. The *ID* in its original form is defined as a logarithm to the base 2 of the doubled distance (movement amplitude) divided by the target width: $ID = \log_2(2A \ / \ W)$. For the mouse movements the more appropriate is the Fitts's *ID* formula proposed by Welford (1960): $ID = \log_2(A \ / \ W + 0.5)$.

In the present experiment, taking the value of a movement amplitude as the distance between the centre of the *START* button and the panel centroid, one can approximate average *ID*s for individual layouts. The lowest Welford *ID* computed in this way was obtained for large items in a layout consisting of two columns and 18 rows (18_02) and amounted to 3.69. The biggest value was acquired for small items in a square layout (06_06) and equalled to 4.73. Given the Fitts's equation parameters provided by MacKenzie (1992) for computer tasks ($MT = 12.8 + 94.7 \cdot ID$), the predicted movement times for these two layouts amounted to 362 and 460 (ms). These values are decidedly smaller than the mean values of acquisition times registered during this investigation which amounted to 1834 and 2359 respectively. The discrepancy clearly shows that the Fitts's law is highly insufficient in accounting for the recorded task completion times and the cognitive aspects should also be taken into account. In this study, the size of the stimulus was the same (36 items) in all experimental conditions so the parameters from the Hick-Hyman's equation cannot be estimated and the search times should be constant. Considering the cognitive and motor tasks carried out together, one could suspect that the most straightforward linear combination of the Fitts's and Hick-Hyman's laws would be suitable. However, such a hypothesis was rejected initially by the research results demonstrated by Hoffmann and Lim (1997) on concurrent manual-decision tasks and later in the studies of Grobelny et al. (2005), and Michalski et al. (2006) in the context of search and click tasks.

The reasoning presented above shows that the issues concerned with the distance to a target seem to be of minor importance especially in relation to the total time necessary of finding and selecting the objects in the current study.

## 3.2. Subjective measures

The obtained mean relative weights calculated across all target sizes showed generally different user behaviour at the beginning of the examination and after accomplishing the tasks. The impact of the panel layout on the users' preferences was meaningful both before and after accomplishing the tasks, however the tendency was opposite. Before the experimental trials more compact layouts were rated lower, which was in contradiction with the objective results. Users expressing their likings after performing 'search and click' tasks favoured decidedly less dispersed layouts and the most vertical panels, though initially strongly preferred, were the worst perceived later. It should also be noted that the significance level for the layout factor before panel operation tasks was equal 0.1, while after the task completion it was distinctly lower (0.002) which may suggest stronger and clearer preference hierarchy occurring after the panel testing.

Although there were no significant differences observed in average 'search and click' times, the registered preferences related to the location variate differed decidedly before the performance tasks. The observed strong preference of left located panels over their right counterparts probably resulted from the users' habits and expectations since the left location of toolbars predominates in many contemporary programs. However, after gaining some experience with the examined layouts, the subjects realized that there is not any substantial difference between these two positions of the examined graphical structures. The presented change in the mean preferences could have also been strengthened or even induced by the simple experimental set-up focused on basic user performance.

The results of the conjoint analysis conducted prior and post the search and click trials showed that subjects treated the layout factor as the most important one. Initially, before the test trials, the discrepancy in relative importances was moderate and amounted to 20%. After the search and click performance the difference significantly rose to 76%. Such a modification of the relative importances provides additional evidence that generally participants changed their initial opinions according to the experience they gained during testing the panels. As the objective results show, the layout factor significantly differentiated the mean acquisition times whereas such a phenomenon was not detected for the location factor. A significant role of the search and click practice session on the subjects' behaviour was additionally confirmed by the choice models results.

The outcomes of the cluster analysis were somewhat surprising. Though the existence of distinct groups of subjects having a similar preference structure before testing the panels were predictable, the three clusters almost equal in size formed after performing the trials were highly unexpected. The participants from the first cluster rated better and better the more and more vertical toolbars. Such a behaviour seems to be particularly strange in the context of the objective results that showed the gradual decrease in average task accomplishment times for the more and more compact configurations. This indicates that there might be some users who are completely insensitive to the practical experience they get. Probably, for the similar reason, the third cluster grouped participants who disliked the square panels, for which the experimental trials were performed the quickest. The existence of those various groups of subjects having a similar preference structure could explain the ambiguous outcomes revealed by the three choice models predicting users' behaviour that were applied within the framework of the conjoint analysis.

## 3.3. Limitations and future research

There are, of course, a number of limitations concerned with this research. First, most of the subjects were experienced computer users, so the results may not be applicable to novice users. No other than 'point and click' technique nor different than square shaped buttons were used in this study. The usage of some other graphical objects than letters and digits could also change the results. Moreover, the data from the demonstrated examination were collected in the laboratory environment and may not fully reflect the real-life situations.

In this research it was indicated that women seem to be less coherent in this type of comparisons, so in subsequent studies it may be recommended to control for the gender effect. The application of the *CR* criterion resulted in the exclusion of many of the gathered results, and some researchers may find the necessity of excluding the inconsistent results inconvenient or troublesome. Therefore, in the next studies, it may be advisable either to decidedly raise the number of participants or to apply one of the existing methods of improving the *CR* values. However, one should bear in mind that the former solution could considerably increase the cost of the study in terms of effort and time. The second proposal, in turn, may be applied two-fold: either force the participants to correct the given preferences to lower the *CR* values or to improve the pairwise comparison matrices artificially by mathematical procedures without the user participation (e.g. Maas et al., 1995).

Given the significant alteration of the subjects' preferences that was revealed in the current research, it is interesting to what degree the change in preferences is stable over time, and across the groups identified by the cluster analysis. The future studies may be focused on these issues.

## 4. Conclusions

There is no doubt that analysing, modelling and determining the real structure of the people's preferences is essential in many areas. The importance of the subjective attitudes towards interactive systems is also gaining more and more attention among the HCI researchers. In the present study, subjective users' opinions towards examined structures considerably changed after testing the panels in relation both to the layout and location factors. As we could see in the current research the users' habits may significantly influence their attitude towards given graphical solution. The practitioners should also be aware that the initial preferences may significantly change during the software usage, and that the potential users are not homogenous with respect to their attitudes towards graphical interfaces.

Apart from the main findings concerned with the subjective issues, this investigation provided some evidence that in terms of the user performance pertaining to simple tasks involving both cognitive processing and visually controlled motor activities, the panel location does not play an important role. Furthermore, it was confirmed that the toolbar layout and target size significantly affect acquisition times.

The application of both subjective and objective measures concurrently in a single study seems to be necessary in many usability researches as they may provide complementary results. Moreover, to the best of the author's knowledge no previous HCI studies employed the Analytic Hierarchy Process framework along with the conjoint and cluster analyses. The inclusion of the AHP method for determining the users' preferences may be helpful especially for the sake of controlling the reliability of the obtained preferences, what might have an impact on conclusions derived from performed analyses. The *CR* values may also indicate that the differences between the examined options are so small or unclear that the subjects may have difficulty in providing consistent ratings. Anyway, this parameter allows for a more thorough and in-depth analysis of the gathered data. Taking advantage of the AHP approach one should also be aware of the major drawback of this tool, namely the rapidly growing number of comparisons along with the increasing number of decision variants.

The other two methods used for the subjective data analysis provided also valuable information and demonstrated various faces of the obtained preferences. The conjoint analysis specified relative importances of the examined independent factors whereas the clustering technique showed that there existed versatile groups of participants sharing the same preference patterns. Each of the applied methods naturally has its advantages and limitations and they should be treated as a set of complementary techniques rather than the mutually exclusive ones. Taking advantage of various approaches will certainly allow for better understanding of the people's preference nature. This, in turn, may help to make appropriate practical and scientific decisions. In summary, the following practical implications for the interface design arise from the current study results:

(1) Generally, the layout of the graphical structures is subjectively much more important to users than their location on the screen. This information may be helpful in specifying the priorities during the interface design process.

(2) The designers should be aware that the users' expectations may not reflect the objective usefulness of the given solution but their initial attitudes may be comparatively quickly changed. Thus, a possible strategy could consist in encouraging the potential users to test and get familiar with the objectively better solution.

(3) When there is no difference in objective measures the decisions regarding the specific solution can be based on the preference examination to conform to the users' habits.

(4) There might be distinct groups of people with various opinions among the users and some of them do not change their minds even after the task performance experience. They should have the possibility of customizing the default design solution.